\documentclass[atlasdraft=false, REPORT=false, UKenglish, coverpage=false, paper=a4, PAPER=true, cernpreprint=true, texmf, orcidlogo]{atlasdoc}
\usepackage[subfigure]{atlaspackage}

\usepackage{atlasbiblatex}

\usepackage{atlasphysics}
\usepackage{atlasjetetmiss}

\graphicspath{{logos/}{figures/}}

\usepackage{ANA-STDM-2023-07-PAPER-defs}

\AtlasTitle{Measurements of Lund subjet multiplicities in 13 TeV proton-proton collisions with the ATLAS detector}
\AtlasAbstract{%
This Letter presents a differential cross-section measurement of Lund subjet multiplicities, suitable for testing current and future parton shower Monte Carlo algorithms.
This measurement is made in dijet events in 140~fb$^{-1}$~of $\sqrt{s}=13$~TeV proton--proton collision data collected with the ATLAS detector at CERN's Large Hadron Collider.
The data are unfolded to account for acceptance and detector-related effects, and are then compared with several Monte Carlo models and to recent resummed analytical calculations.
The experimental precision achieved in the measurement allows tests of higher-order effects in QCD predictions.
Most predictions fail to accurately describe the measured data, particularly at large values of jet transverse momentum accessible at the Large Hadron Collider, indicating the measurement's utility as an input to future parton shower developments and other studies probing fundamental properties of QCD and the production of hadronic final states up to the TeV-scale.
}

\AtlasRefCode{STDM-2023-07}

\PreprintIdNumber{CERN-EP-2024-029}

\AtlasDate{\today}

\arXivId{2402.13052}

\HepDataRecord{ins2759899}

\AtlasJournalRef{\PLB 859 (2024) 139090 }
\AtlasDOI{10.1016/j.physletb.2024.139090}

\DeclareBoolOption[true]{LANGEDIT}
\LEcontact{Chip~Brock}

\hypersetup{pdftitle={ATLAS document},pdfauthor={The ATLAS Collaboration}}

\begin{document}

\maketitle

% The next lines are included from the .//sections/body.tex input file
%
\newcommand{\AtlasCoordFootnote}{%
ATLAS uses a right-handed coordinate system with its origin at the nominal interaction point (IP) in the centre of the detector and the \(z\)-axis along the beam pipe. The \(x\)-axis points from the IP to the centre of the LHC ring, and the \(y\)-axis points upwards. Polar coordinates \((r,\phi)\) are used in the transverse plane, \(\phi\) being the azimuthal angle around the \(z\)-axis. The pseudorapidity is defined in terms of the polar angle \(\theta\) as \(\eta = -\ln \tan(\theta/2)\) and is equal to the rapidity $ y = \frac{1}{2} \ln \left( \frac{E + p_z c}{E - p_z c} \right) $ in the relativistic limit. Angular distance is measured in units of \(\Delta R \equiv \sqrt{(\Delta y)^{2} + (\Delta\phi)^{2}}\).}

\section{Introduction}

At particle colliders such as the Large Hadron Collider (LHC) at CERN, the high-energy quarks and gluons that are produced in hard-scattering (HS) processes fragment and hadronise, producing collimated \emph{jets} of hadrons in the final state~\cite{Salam:2010nqg}.
Jets are complex objects; and jet substructure (JSS) probes a wide range of energy scales, providing a multi-faceted setting for tests of quantum chromodynamics (QCD) in the colinear limit~\cite{Marzani:2019hun}.
Parton Shower Monte Carlo (PSMC) programs  model the evolution of quarks and gluons produced in the HS into the low-energy hadrons that are observed experimentally, and are among the most widely-used theoretical tools in particle physics~\cite{Buckley:2011ms}.
Modern PSMCs can adequately describe the bulk of LHC data, but predictions from different algorithms can disagree significantly for certain topologies or processes.
Their formal accuracy is also limited in cases where subsequent emissions within the shower have commensurate energies or angles~\cite{FerrarioRavasio:2023kyg}: in such configurations, emissions beyond the initial one must be considered to obtain an accurate result.
This lack of accuracy affects the precision of data analysis, either through \emph{ad hoc} comparisons of discrepant algorithms (\emph{e.g.} Refs.~\cite{ATL-PHYS-PUB-2018-009,STDM-2017-27,JETM-2018-05,CMS-JME-13-004}) or due to increased reliance on particle-level correlations in simulation (\emph{e.g.} with supervised Machine Learning techniques~\cite{Larkoski:2017jix}).

There has been progress toward improved understanding of higher-order QCD effects in PSMCs and the development of more accurate algorithms~\cite{Bewick:2019rbu,Dasgupta:2020fwr,Forshaw:2020wrq,Nagy:2020dvz,Nagy:2020rmk,Bewick:2021nhc,vanBeekveld:2022zhl,vanBeekveld:2022ukn,Herren:2022jej,vanBeekveld:2023chs,FerrarioRavasio:2023kyg}.
A necessary advancement for this effort is the incorporation of `double-soft' splittings~\cite{Dokshitzer:1990yi,Catani:1999ss}, which describe the emission of two soft gluons or a quark-antiquark pair, beyond tree-level in QCD~\cite{Zhu:2020ftr,Catani:2021kcy,Czakon:2022dwk}.
These processes have been implemented in some existing PSMC programs~\cite{Jadach:2011kc,Hartgring:2013jma,Jadach:2013dfd,Jadach:2016zgk,Li:2016yez,Hoche:2017iem,Hoche:2017hno,Dulat:2018vuy,Campbell:2021svd,Gellersen:2021eci}, and are now being implemented in PSMCs with higher logarithmic accuracy~\cite{FerrarioRavasio:2023kyg}.
The `Lund' subjet multiplicity, or `Lund multiplicity,' is a JSS observable used to test for the inclusion of double-soft splittings~\cite{FerrarioRavasio:2023kyg}.
For this purpose, it was calculated with analytical resummation at next-to-next-to-double-logarithmic (NNDL) accuracy in QCD in Refs.~\cite{Medves:2022ccw,Medves:2022uii}.
These predictions were used to assess the implementation of these higher-order contributions within the general-purpose \textsc{PanGlobal} PSMC~\cite{Dasgupta:2020fwr}.

The Lund multiplicity counts the number of subjets in a jet's angle-ordered clustering history~\cite{Medves:2022ccw} above a specified requirement on the subjet's transvese momentum relative to the jet core.
This angular-ordered clustering history is obtained by reclustering the jet's constituents with the Cambridge-Aachen (C/A) algorithm~\cite{Dokshitzer:1997in,Wobisch:1998wt}\footnote{\AtlasCoordFootnote}.
The branches of the clustering history with lower and higher relative transverse momentum at each step are respectively referred to as the `emission' and `core' of the jet, such that the transverse momentum of an emission relative to the jet core may be written as
\begin{equation*}
\kt = \pt^{\text{emission}} \cdot \DeltaR(p^\text{emission}, p^\text{core}).
\end{equation*}

\begin{figure}
\centering
\includegraphics[width=0.49\textwidth]{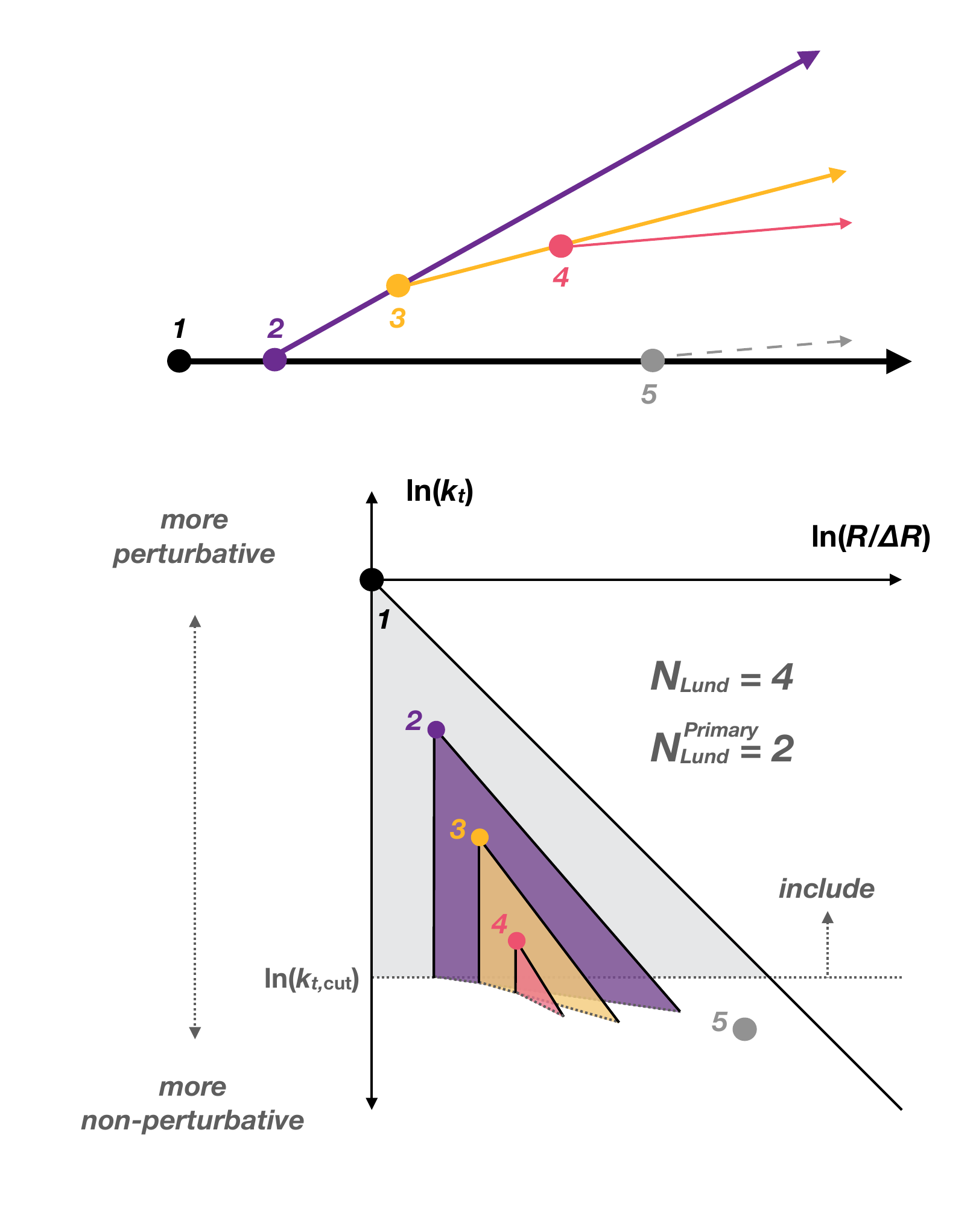}
\caption{
Schematic representation of a jet's angle-ordered clustering history and the calculation of \nlund and \nprimary.
The emission labelled `5,' drawn with a dashed line, fails the \kt requirement and is not counted in the multiplicity calculation.
}
\label{fig:cartoon}
\end{figure}

The Lund multiplicity is built from the picture of JSS based on the Lund jet plane~\cite{Andersson:1988gp,Dreyer:2018nbf}, which has proved to be a powerful experimental tool: the Primary Lund jet plane was measured by the ATLAS, ALICE and CMS collaborations~\cite{STDM-2018-57,ALICE:2021yet,CMS-SMP-22-007}, was used to develop new identification and calibration algorithms for boosted hadronically-decaying resonances~\cite{ATL-PHYS-PUB-2023-017,CMS-DP-2023-046}, and was used by the ALICE Collaboration to observe the dead-cone effect in QCD~\cite{ALICE:2021aqk}.
The Lund multiplicity may be counted either in the full, fractal Lund jet plane (\nlund) by following each branch of the clustering history, or along only the primary clustering sequence (\nprimary).
These schemes provide observables with different levels of sensitivity to perturbative and non-perturbative effects.
A schematic representation of this procedure for a jet with $\nlund=4$ and $\nprimary=2$ is provided in Figure~\ref{fig:cartoon}.
No measurement of the Lund multiplicity has been performed to-date.
However, the related Cambridge multiplicity has an alternative definition of \kt that is suitable for $e^+e^-$ annihilation, and was measured at LEP by the OPAL Collaboration~\cite{JADE:1999zar}.
New, precise measurements of observables used to assess progress in PSMC development are also an important component of the collective effort to understand QCD in the colinear limit at the level necessary to fully exploit the LHC data~\cite{ATL-PHYS-PUB-2022-018}.

This Letter presents a differential cross-section measurement of the Lund subjet multiplicity in dijet events using 140~\ifb of $\sqrt{s}=13~\TeV$ proton--proton ($pp$) collision data collected by the ATLAS detector during Run 2.
The \nlund and \nprimary distributions are measured differentially in up to five bins of jet \pt between 300~\GeV--4500~\GeV, and in relative-rapidity bins that separate in each event the more-forward and more-central of the two selected jets.
These distributions are corrected for detector effects using a regularized, iterative unfolding procedure~\cite{DAgostini:1994fjx}.
\nlund and \nprimary are measured for eight different emission \kt requirements (where \kt is required to be above 0.5~\GeV, 1~\GeV, 2~\GeV, 5~\GeV, 10~\GeV, 20~\GeV, 50~\GeV and 100~\GeV), resulting in sixteen differential measurements.
The measured differential cross-sections as a function of \nlund and \nprimary, along with their their extracted mean values, are compared with state-of-the-art PSMCs.
The average value of \nlund is also compared with the recent next-to-leading-order (NLO) plus NNDL analytic prediction from Ref.~\cite{Medves:2022uii}.
Selected results are presented in the body of this Letter, but the complete differential measurement may be found as Supplemental Material, and digitally on the HepData platform~\cite{hepdata.138878,Maguire:2017ypu}.

\section{The ATLAS Detector}

The ATLAS detector~\cite{PERF-2007-01, ATLAS-TDR-2010-19, PIX-2018-001} is a general-purpose particle detector which provides nearly $4\pi$ coverage in solid angle.
The inner tracking detector (ID) is inside a 2~T magnetic field and measures charged-particle trajectories up to $|\eta|=2.5$.  The innermost component of the ID is a pixelated silicon detector with fine granularity capable of resolving ambiguities inside the dense hit environment of jet cores~\cite{PERF-2015-08}, surrounded by silicon microstrip and transition radiation detectors.
Beyond the ID are electromagnetic and hadronic calorimeters, from which topologically connected clusters of cells~\cite{PERF-2014-07} are formed into particle-flow objects~\cite{PERF-2015-09} using extrapolated measurements of charged ID particles.
A muon spectrometer incorporating three superconducting air-core toroidal magnets systems surrounds the calorimetry systems.
An extensive software suite is used in the reconstruction and analysis of real and simulated data, detector operations, and in the ATLAS trigger and data acquisition systems~\cite{ATL-SOFT-PUB-2021-001}.

\section{Reconstruction and event selection}

Data passing the single-jet trigger selection~\cite{TRIG-2016-01,TRIG-2012-01} are recorded, and must satisfy the quality criteria described in Ref.~\cite{DAPR-2018-01}.
The thresholds of these triggers varied depending on the data-taking period during \RunTwo, and some triggers are prescaled during data-taking.
All triggers used to select data for this analysis are fully efficient in the regions where they are used.
Events in data are reweighted by the appropriate prescale factor to recover a smoothly falling jet \pt spectrum in the analysis.

Jets are reconstructed from particle-flow objects~\cite{PERF-2015-09} using the \akt algorithm with radius parameter $R=0.4$~\cite{Cacciari:2008gp,Fastjet}.
Jets are calibrated such that the average detector-level jet energy scale (JES) matches that of the corresponding particle-level jets~\cite{JETM-2018-05}.
Both leading and subleading jets must have at least $\pt>120~\GeV$ and be within the ID acceptance ($|y_{\text{jet}}|<2.1$).
They must also satisfy a dijet balance requirement of
\begin{equation*}
\pt^\text{leading} < 1.5\times \pt^\text{subleading},
\end{equation*}
suppressing possible background contributions and simplifying interpretations of the final state in terms of a \TwoToTwo process.
Only the leading and subleading jets in selected events are included in this measurement.

ID tracks are used for this measurement to precisely reconstruct the trajectories of individual particles.
This approach follows methodology of earlier measurements by ATLAS~\cite{STDM-2017-04,STDM-2017-33,STDM-2018-57} and CMS~\cite{CMS-SMP-20-010, CMS:2023lpp}, providing good experimental resolution of colinear splittings within the dense jet core where relevant angular scales are much smaller than the ATLAS calorimeter granularity.
Tracks are required to have $\pt > 500~\MeV$ and to be matched to the reconstructed primary vertex with the largest sum of track $\pt^2$ in the event~\cite{ATL-PHYS-PUB-2015-051}.
All tracks within $\DeltaR=0.4$ of selected \akt jets are reclustered using the C/A algorithm.
The clustering history of the jet is then iteratively declustered to obtain the Lund multiplicities, counting the splittings within the C/A sequence that satisfy the emission \kt requirement.
Once a splitting is encountered that fails the emission \kt requirement, no additional splittings are considered for that branch.
To account for contributions from neutral particles, the emission \kt reconstructed from stable charged particles is rescaled by the ratio of the jet \pt measured from stable charged and neutral constituents over the value of the jet \pt measured only from its charged constituents:
\begin{equation*}
\kt = ( \pt^{\text{all}} / \pt^{\text{charged}} ) \kt^{\text{charged}}.
\end{equation*}
As isospin is an approximate symmetry of the strong force, once the \kt value has been rescaled, the average difference between multiplicities constructed using all interacting particles and only stable charged hadrons is expected to be smaller~\cite{STDM-2018-57,Lifson:2020gua,CMS:2023lpp}.
All multiplicities presented in this analysis, at detector- and particle-level, are computed in this way.

\section{Simulated event samples}

Dijet events were simulated to perform the unfolding and to compare with the measured cross-sections.

\PYTHIA[8.230]~\cite{Sjostrand:2014zea,Sjostrand:2006za} is used as the nominal MC generator for this analysis, and is also referred to here as the `nominal' simulation.
Samples of $\TwoToTwo$ dijet events were simulated using the A14 tune~\cite{ATL-PHYS-PUB-2014-021}, the Lund string hadronisation model and the \NNPDF[2.3lo]~\cite{Ball:2012cx} leading-order (LO) parton distribution function (PDF) set.
The \PYTHIA parton shower (PS) algorithm uses a dipole-style \pt-ordered evolution, and its renormalisation and factorisation scales were set to the geometric mean of the squared transverse masses of the outgoing particles.
\EVTGEN~\cite{Lange:2001uf} was used to model decays of heavy-flavour hadrons.

Two sets of \SHERPA[2.2.5]~\cite{Gleisberg:2008ta} dijet events were simulated with the default AHADIC cluster hadronisation model~\cite{Webber:1983if} or with the \SHERPA interface to the Lund string hadronisation model as implemented in \PYTHIA[6.4], and its decay tables.
These samples include LO matrix element calculations for \TwoToTwo processes, and use the \SHERPA parton shower algorithm based on Catani--Seymour dipole subtraction~\cite{Schumann:2007mg}.
The \CT[14nnlo] next-to-next-to-leading-order (NNLO) PDF~\cite{Dulat:2015mca} set is used for matrix element calculations and \CT[10] is used for multi-parton interactions (MPI)~\cite{Lai:2010vv}.

Two additional sets of \SHERPA[2.2.11] samples were generated using \TwoToTwo MEs and the CT10 PDF set, with either the default \pt-ordered~\cite{Schumann:2007mg} or the alternative \DIRE~\cite{Hoche:2015sya} PS algorithms.
These samples use an updated set of tuned parameters for the cluster hadronisation model~\cite{Chahal:2022rid}.
The \DIRE sample is of particular interest in the context of this analysis, as it incorporates some aspects of higher-order splitting functions in QCD~\cite{Dulat:2018vuy}, although not at the level of NLL accuracy~\cite{Dasgupta:2020fwr}.

A sample of \HERWIG[7.1.3]~\cite{Bahr:2008pv,Bellm:2015jjp,Bellm:2017bvx} multijet events was generated with the \MMHT[nlo] PDF set~\cite{Harland-Lang:2014zoa}, default cluster hadronisation model and the default angle-ordered PS.
This sample models \TwoToTwo~matrix elements with NLO accuracy and~\TwoToThree~matrix elements with LO accuracy.
The parton shower was matched to the matrix element calculation using the \MCatNLO matching scheme~\cite{Lonnblad:2001iq,Lonnblad:2011xx}, and the \pt of the leading jet is taken as the renormalisation scale.

A sample of dijet events with NLO matrix element accuracy were produced with \POWHEG[v2]~\cite{Nason:2004rx,Frixione:2007vw,Alioli:2010xa} using the dijet process implemented in \POWHEGBOX[v2]~\cite{Alioli:2010xd}, matched to the \PYTHIA[8] parton showers configured similarly to the corresponding sample described above.
The renormalisation and factorisation scales in these samples were set to the \pt of the underlying Born-level configuration.

An additional \SHERPA[3] prediction was provided by the authors of Ref.~\cite{Herren:2022jej} using LO \TwoToTwo MEs and the \textsc{Alaric} PS~\cite{Herren:2022jej,Assi:2023rbu}, extended to cover initial-state radiation.
This sample uses CT14NNLO PDFs and Lund string hadronisation \emph{via} an interface to \PYTHIA[8].
The \textsc{Alaric} PS is the successor to the \DIRE algorithm, and replaces explicit angle-ordering with a differentially-accurate simulation of the dependence of soft gluon splitting functions on azimuthal angles that is NLL-accurate.

All generated events used in the unfolding and its associated uncertainties were passed through a full detector simulation~\cite{SOFT-2010-01} based on \GEANT~\cite{Agostinelli:2002hh} and overlaid with simulated minimum-bias interactions generated using \PYTHIA[8]~with the A3 tune~\cite{ATL-PHYS-PUB-2016-017} and \NNPDF[2.3lo] PDF set~\cite{Ball:2012cx} to represent the effect of multiple $pp$ interactions in the same and neighboring bunch crossings (`pile-up').
The distribution of the average number of pile-up interactions in simulation was reweighted during data analysis to match that observed in \RunTwo data.
Additional details of the MC samples used in this measurement may be found in Ref.~\cite{ATL-PHYS-PUB-2019-017}.

\section{Unfolding procedure}

Data are unfolded to correct for detector biases, resolution, and acceptance effects by applying iterative Bayesian unfolding~\cite{DAgostini:1994fjx}, as implemented in \textsc{RooUnfold}~\cite{Adye:2011gm}, with four iterations.
The number of iterations was selected based on a comparison of the modelling, statistical and data-driven non-closure uncertainties (described later, in Section~\ref{sec:unc}) to minimise the total uncertainty in the measured cross-sections.
The nominal \PYTHIA MC sample is used to unfold the data.
Unfolding corrects the multiplicities reconstructed from detector-level objects to charged-particle level.
Charged particles and jets at the particle level are defined similarly to those at the detector level.
Particle-level jets are reconstructed using all stable particles with a lifetime $\tau$ in the laboratory frame such that $c\tau > 10$~mm and the same \akt algorithm.
The same kinematic requirements as for detector-level jets are imposed on these jets; charged particles with $\pt > 500$~\MeV\ within $\Delta R = 0.4$ of the cores of particle-level jets are used to calculate \nlund and \nprimary, and the emission \kt is also scaled by the ratio of the jet \pt computed with charged and neutral particles to that computed with only charged particles.

Jets at detector-level and particle-level that are uniquely matched based on their geometrical proximity (\emph{via} a matching based on $\Delta R$) are used to construct a \emph{response matrix}, mapping migrations of the measured observables between detector- and particle-level.
Unmatched jets are accounted for with purity and efficiency corrections,  respectively applied before and after the regularized inversion of the response matrix.
For each multiplicity observable measured in this analysis, this matrix is constructed differentially in the observable itself, and also in bins of jet \pt and two bins of relative jet rapidity ordering that contain the more-central and more-forward jet of each dijet system.
This separation in rapidity is performed to allow the unfolding to properly account for underlying differences between the jet energy scale and fragmentation modelling caused by the different fractions of quark- and gluon-initiated jets in the more-central and more-forward rapidity bins.
An additional \pt bin from 240--300~\GeV is included in the unfolding to reduce the size of the purity and efficiency corrections in the fiducial region, such that they are below 10\%.
Unfolded distributions are normalized to the number of selected jets in a given \pt and relative-rapidity bin, rendering the analysis insensitive to the total jet cross-section and therefore enabling a more precise measurement of the multiplicity distribution's shape.

Since the multiplicity is an integer observable: \nlund and \nprimary distributions are unfolded using unit-width bins to simplify the calculation of the average values for different emission \kt requirements.
After unfolding, the distributions are rebinned (by grouping unit-width bins) based on the detector resolution to obtain the measured differential cross-sections.
Rebinning regularizes the reported distributions, ensuring that they are insensitive to sub-detector-resolution effects and unfolding biases from the steeply-falling distribution in wide bins~\cite{Stanley:2021xce}.
The mean values of \AvgNLund and \AvgNPrimary are computed from the finely-binned distributions; averaging the distribution is another form of regularization.
Studies using alternative methods, such as unfolding with the coarse measurement bins, or fitting the coarsely-binned distributions and applying a binning correction to determine the mean, showed that no significant differences in either the measured values or their uncertainties occur from using the rebinning approach selected for this analysis.

\section{Uncertainties}\label{sec:unc}

Experimental systematic uncertainties are evaluated by independently propagating variations from each source through unfolding (defining new purity and efficiency corrections and a new response matrix), then comparing the varied and nominal results.
JES uncertainties~\cite{PERF-2016-04} cause migrations between jet \pt bins and multiplicity bins due to the charged-to-all particles scaling applied to the emission \kt, typically amounting to $2-4\%$.
Track reconstruction uncertainties are considered by modifying the measured track \pt or removing them completely~\cite{PERF-2015-08,PERF-2015-07}.
Additional uncertainties are also included to account for differences in the tracking efficiency in the dense core of jets~\cite{PERF-2015-08}.
These uncertainties are typically below $2\%$ but increase for smaller \kt requirements.
Other experimental uncertainties related to the modelling of pile-up and the stability of the measurement across data-taking periods are negligible. %
A data-driven non-closure uncertainty that accounts in-part for possible bias due to regularisation in the unfolding procedure (labelled `Unfolding' in Figures~\ref{fig:results:xs}~and~\ref{fig:results:avg}) is determined by unfolding the simulated detector-level distribution following a reweighting based on a comparison of this distribution with the data~\cite{Malaescu:2009dm}; this uncertainty is smaller than $2\%$ and never dominant.

The choice of the parton shower and hadronisation models used for the unfolding procedure can affect the efficiency/purity corrections, response matrix, and unfolding prior.
These contributions are estimated by unfolding with response matrices constructed with the \SHERPA[2.2.11] AHADIC and \HERWIG[7.1] angle-ordered samples.
The component of this uncertainty related to modelling of the unfolding prior is omitted in this procedure, as it is estimated in a data-driven way when estimating the non-closure uncertainty.
As the correlation between these uncertainty sources is unknown, the effect of changing the nominal model on each is assessed separately, then added in quadrature for each MC.
An envelope constructed from the uncertainties determined using these alternative samples is taken as the final uncertainty; the contribution from \HERWIG is usually the largest.
To avoid double-counting of modelling effects related to the JES, which are accounted for separately, the JES of these samples is re-calibrated to match that of the nominal \PYTHIA sample.
This correction is applied separately for quark- and gluon-initiated jets in simulation, as labelled by the highest-energy parton that is associated to the particle-level jet~\cite{PERF-2013-02}.
The uncertainty due to the choice of nominal MC model ranges from $2-10\%$, and is the largest source of uncertainty in the measurement for both the differential cross-sections and extracted average multiplicities.

Statistical uncertainties due to the finite size of the data and MC samples are evaluated with Poissonian pseudo-experiments~\cite{ATL-PHYS-PUB-2021-011} respectively for the data prior or response matrix, purity and efficiency factors during unfolding.
They are less than 1\% throughout the measurement except in the tails of multiplicities.

\section{Results}

\begin{figure}[htbp]
\centering
\subfigure[]{\includegraphics[width=0.38\textwidth, trim=4 8 15 8, clip]{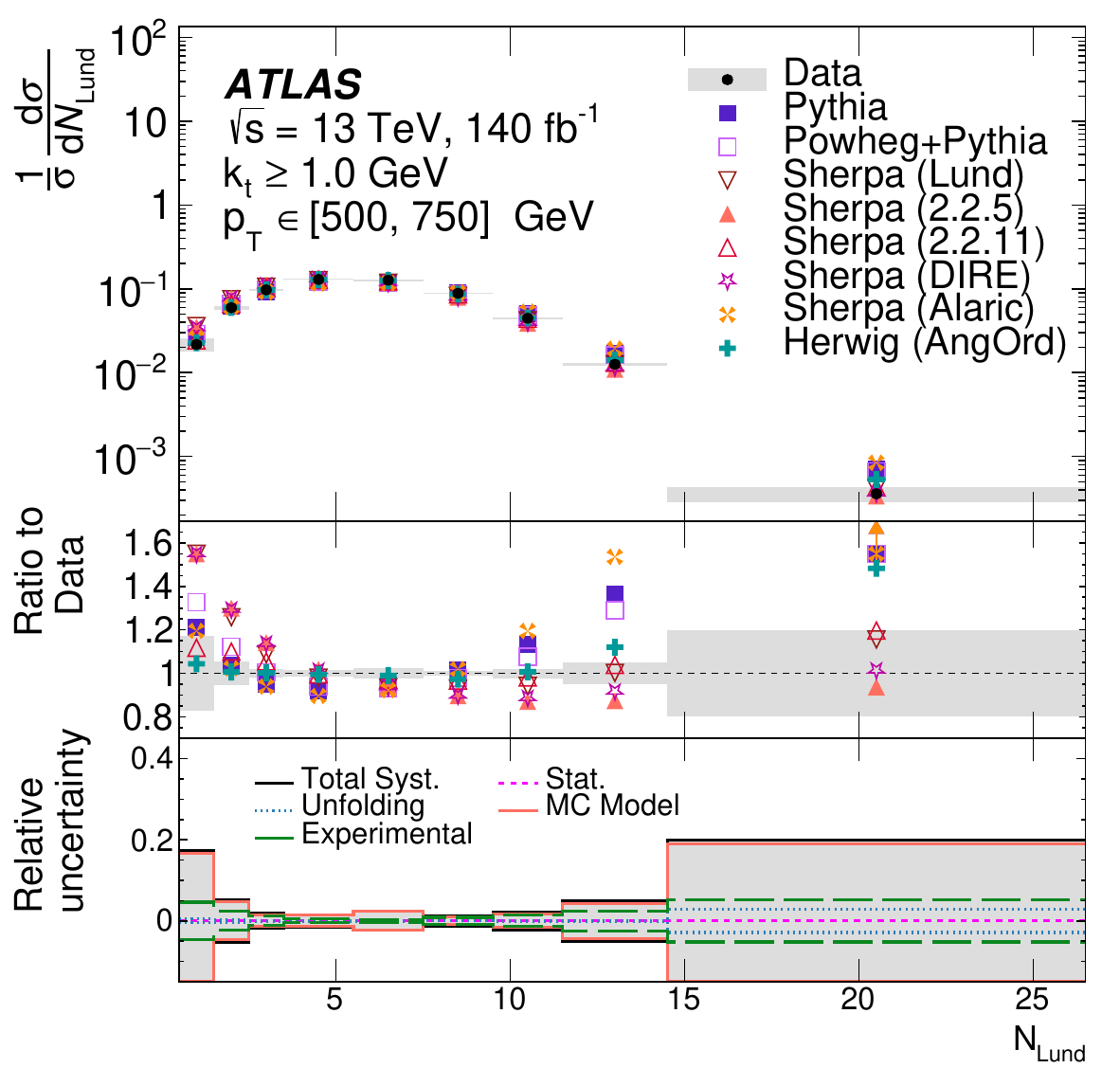}\label{fig:results:xs:a}}
\subfigure[]{\includegraphics[width=0.38\textwidth, trim=4 8 15 8, clip]{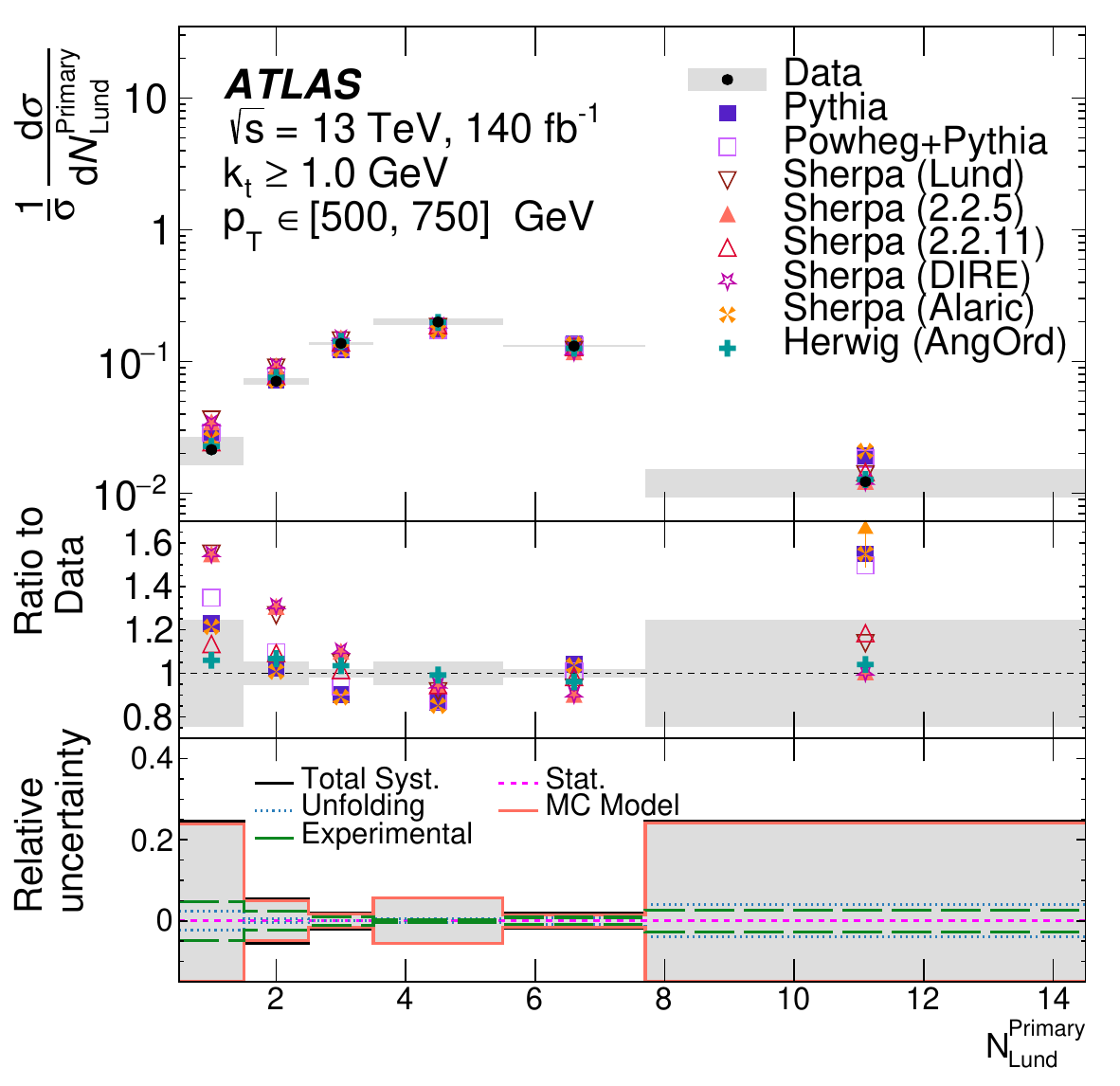}\label{fig:results:xs:b}}
\subfigure[]{\includegraphics[width=0.38\textwidth, trim=4 8 15 8, clip]{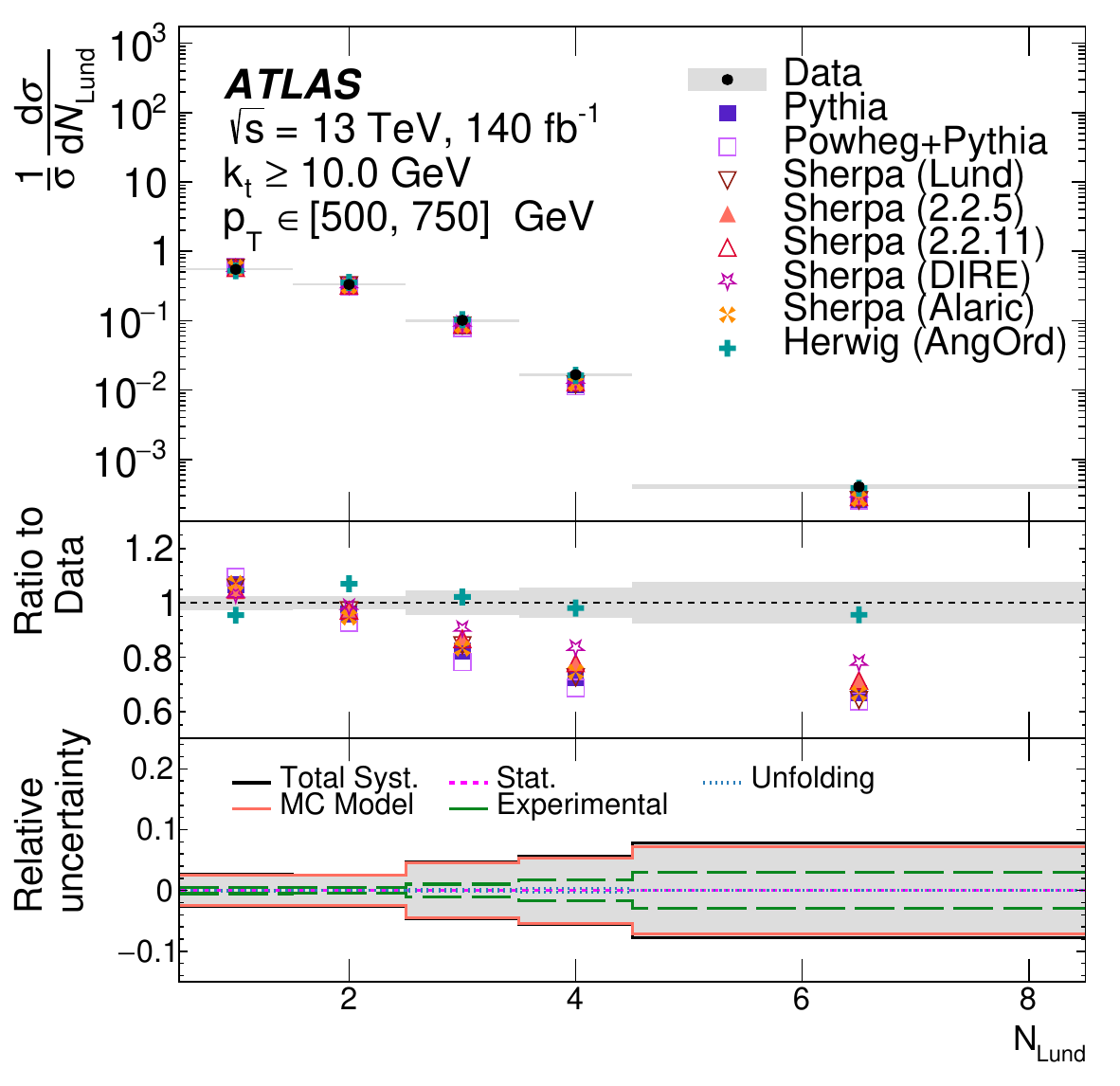}\label{fig:results:xs:c}}
\subfigure[]{\includegraphics[width=0.38\textwidth, trim=4 8 15 8, clip]{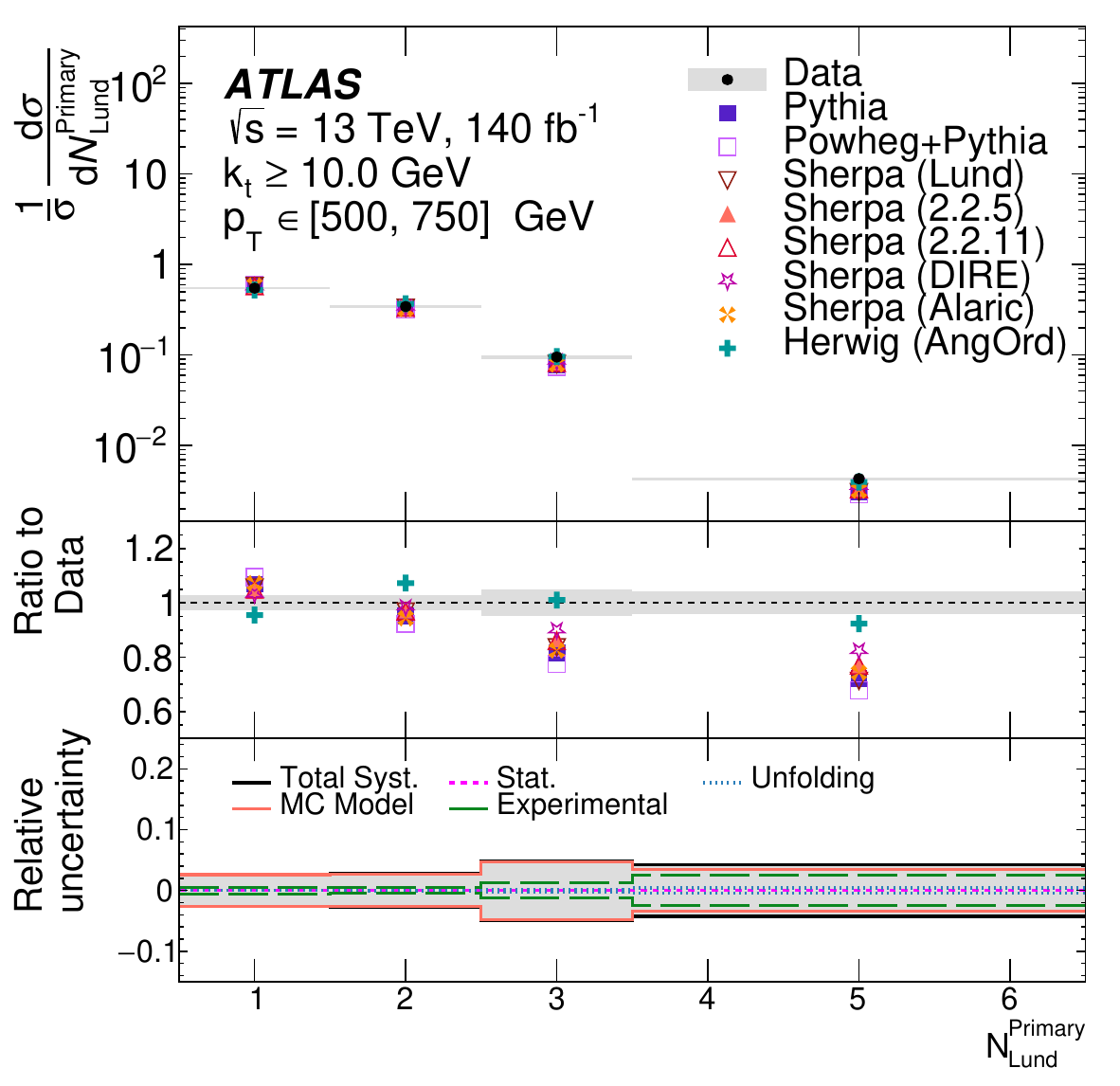}\label{fig:results:xs:d}}
\subfigure[]{\includegraphics[width=0.38\textwidth, trim=4 8 15 8, clip]{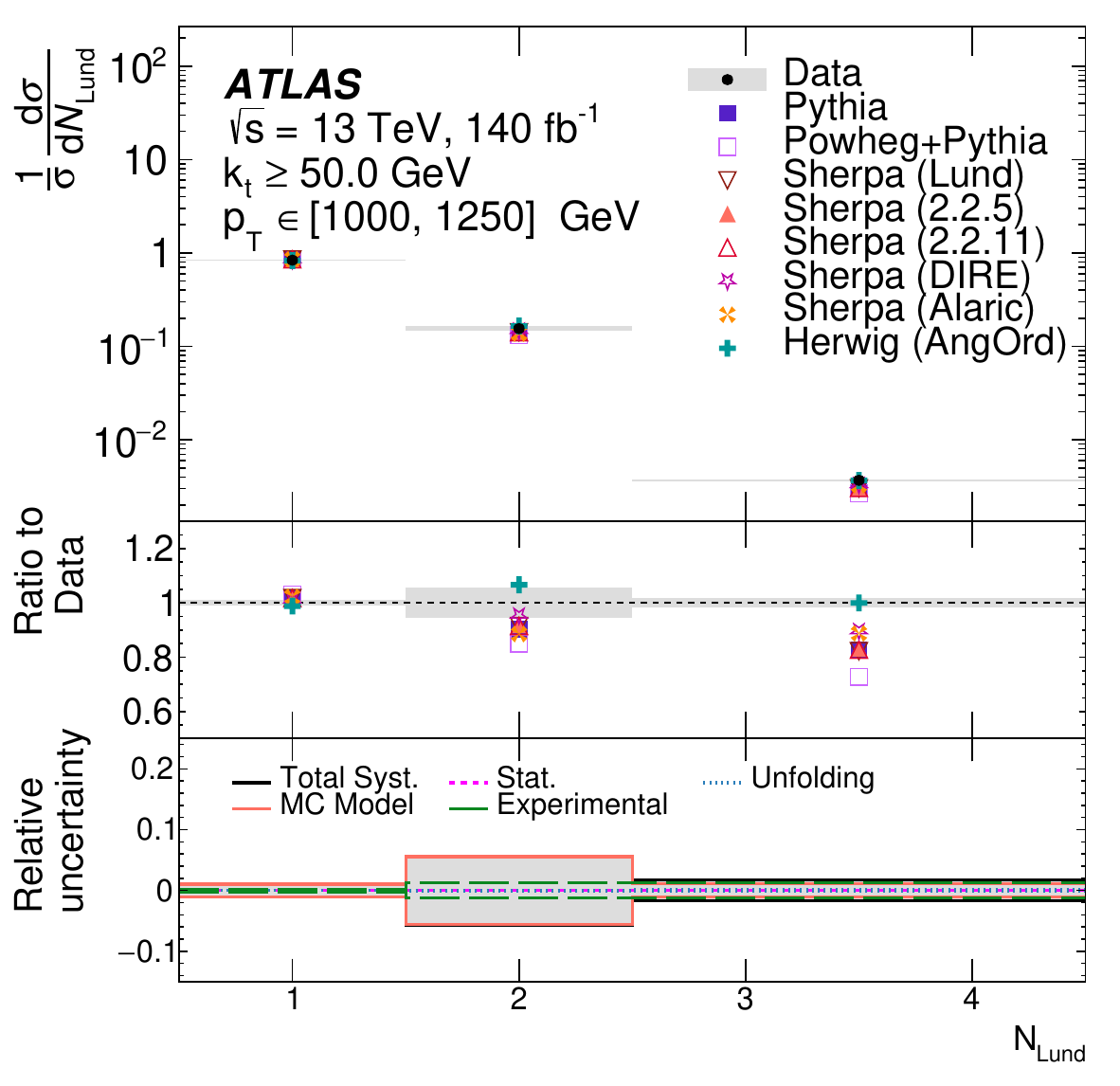}\label{fig:results:xs:e}}
\subfigure[]{\includegraphics[width=0.38\textwidth, trim=4 8 15 8, clip]{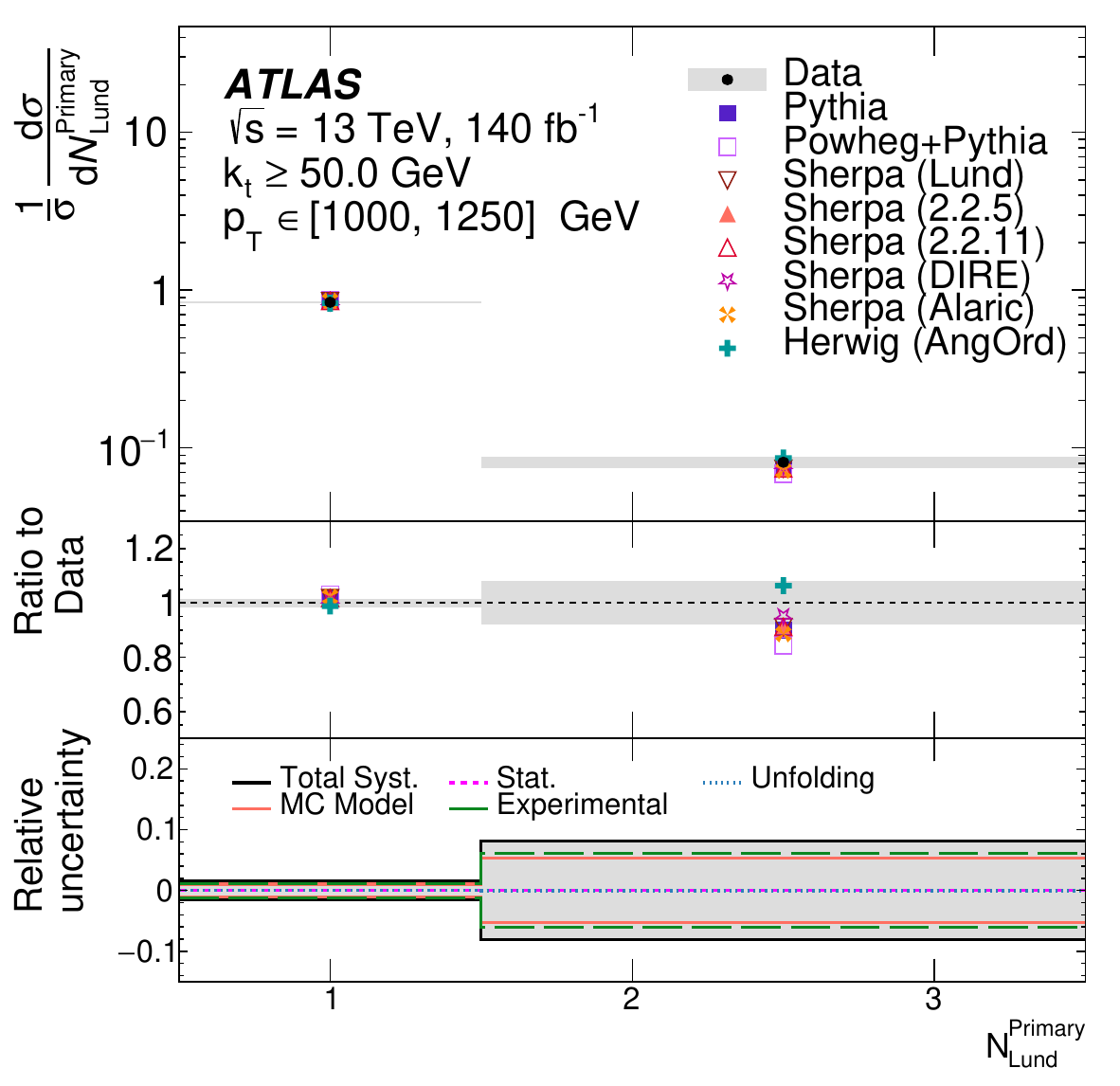}\label{fig:results:xs:f}}
\caption{
A representative measured differential cross-section of (a,c,e) \nlund and (b,d,f) \nprimary for an emission \kt requirement of (a,b) 1~\GeV, (c,d) 10~\GeV and (e,f) 50~\GeV.
The data are shown in an inclusive bin of jet rapidity and a bin of jet \pt between either (a-d) 500-750~\GeV or (e,f) 1000-1250~\GeV.
The unfolded data are compared with several MC predictions, and the total uncertainty on the data is indicated by a shaded grey region.
The middle panel shows a ratio of the predictions to the measured data, and the bottom panel summarizes the various systematic uncertainties in each bin.
}
\label{fig:results:xs}
\end{figure}

A representative selection of the measured differential cross-section as a function of \nlund and \nprimary is presented in Figure~\ref{fig:results:xs}, where the unfolded data are compared with several PSMCs.
Most of the studied MC models do not describe the shape of the measured data well, showing two patterns of disagreement from small to large multiplicity values.
For multiplicities with small \kt requirements, significant data-to-MC disagreements are generally observed at both low and high multiplicities.
The total uncertainty on the measured differential cross-sections decreases with increasing \kt, and depends on the jet \pt bin.
It is typically between $2-5\%$ in the bulk and $5-30\%$ in the tails of the multiplicity distributions.

The \HERWIG angle-ordered PS provides the best overall description of both of the observables, once the emission $\kt$ is perturbative ($\geq 1~\GeV$) and for all configurations in the bulk of the multiplicity distributions.
For smaller $\kt$ requirements ($\kt\leq1~\GeV$), the \SHERPA[2.2.11] and DIRE setups do best in the high-multiplicity tail, where non-perturbative contributions can be significant (Figures~\ref{fig:results:xs:a} and~\ref{fig:results:xs:b}).
With larger $\kt$ requirements, most models show a similar decreasing pattern (Figures~\ref{fig:results:xs:c} and~\ref{fig:results:xs:d}).
In these cases, the best performing model tends to change from the \SHERPA samples to the \HERWIG sample as the multiplicity increases.
The \SHERPA sample using the \ALARIC PS is found to behave similarly to the other \SHERPA models in this region, where non-perturbative contributions are not significant.

Even for the largest \kt requirements that can be best-probed in high-\pt jets, such as the $\kt\geq50~\GeV$ results from the bin of jet \pt between $1000~\GeV$ and $1250~\GeV$ shown in Figure~\ref{fig:results:xs:e} and~\ref{fig:results:xs:f}, the \HERWIG sample performs best.

\begin{figure}[htbp]
\centering
\subfigure[]{\includegraphics[width=0.37\textwidth, trim=5 9 15 5, clip]{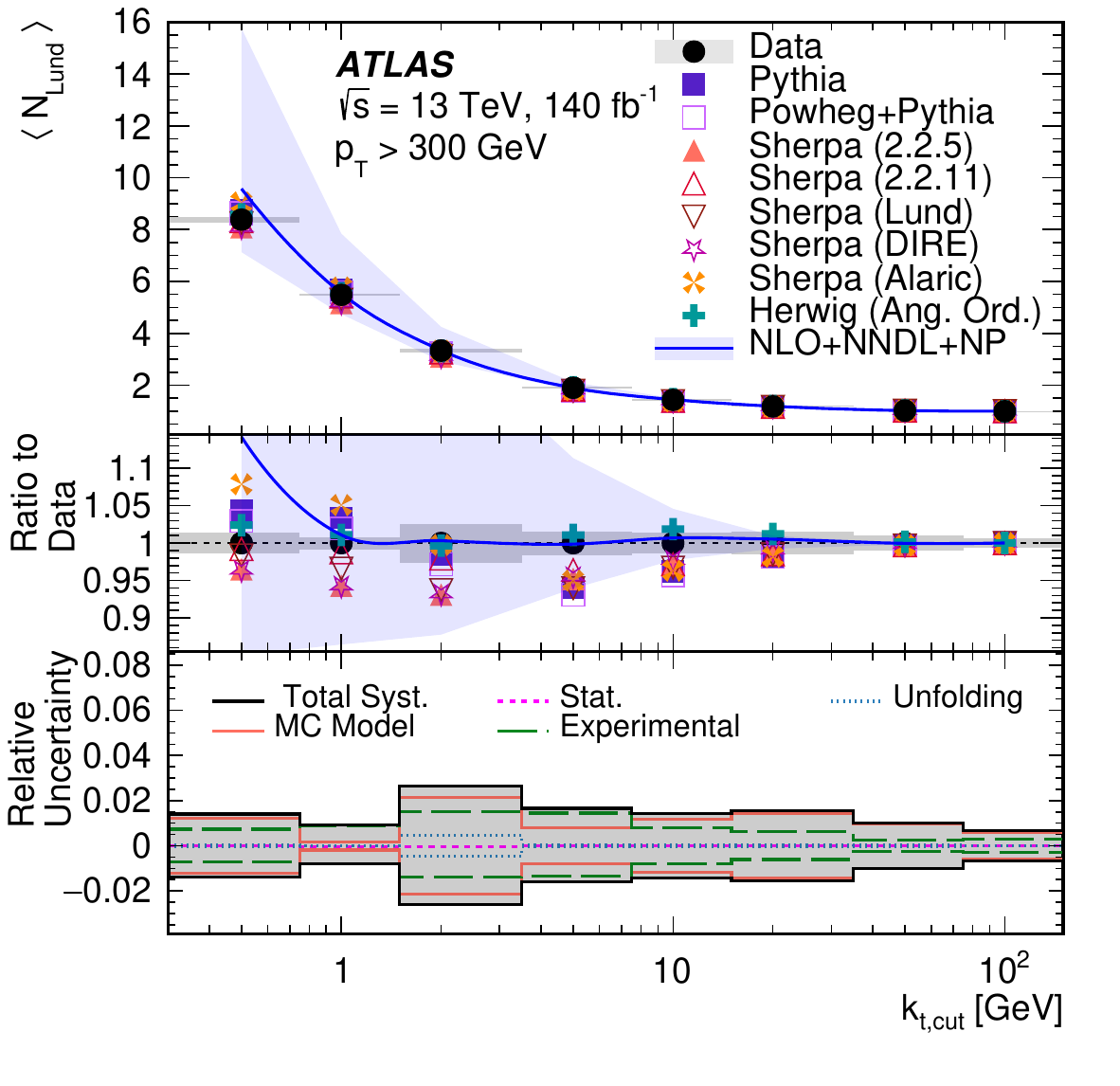}\label{fig:results:avg:a}}
\subfigure[]{\includegraphics[width=0.37\textwidth, trim=5 9 15 5, clip]{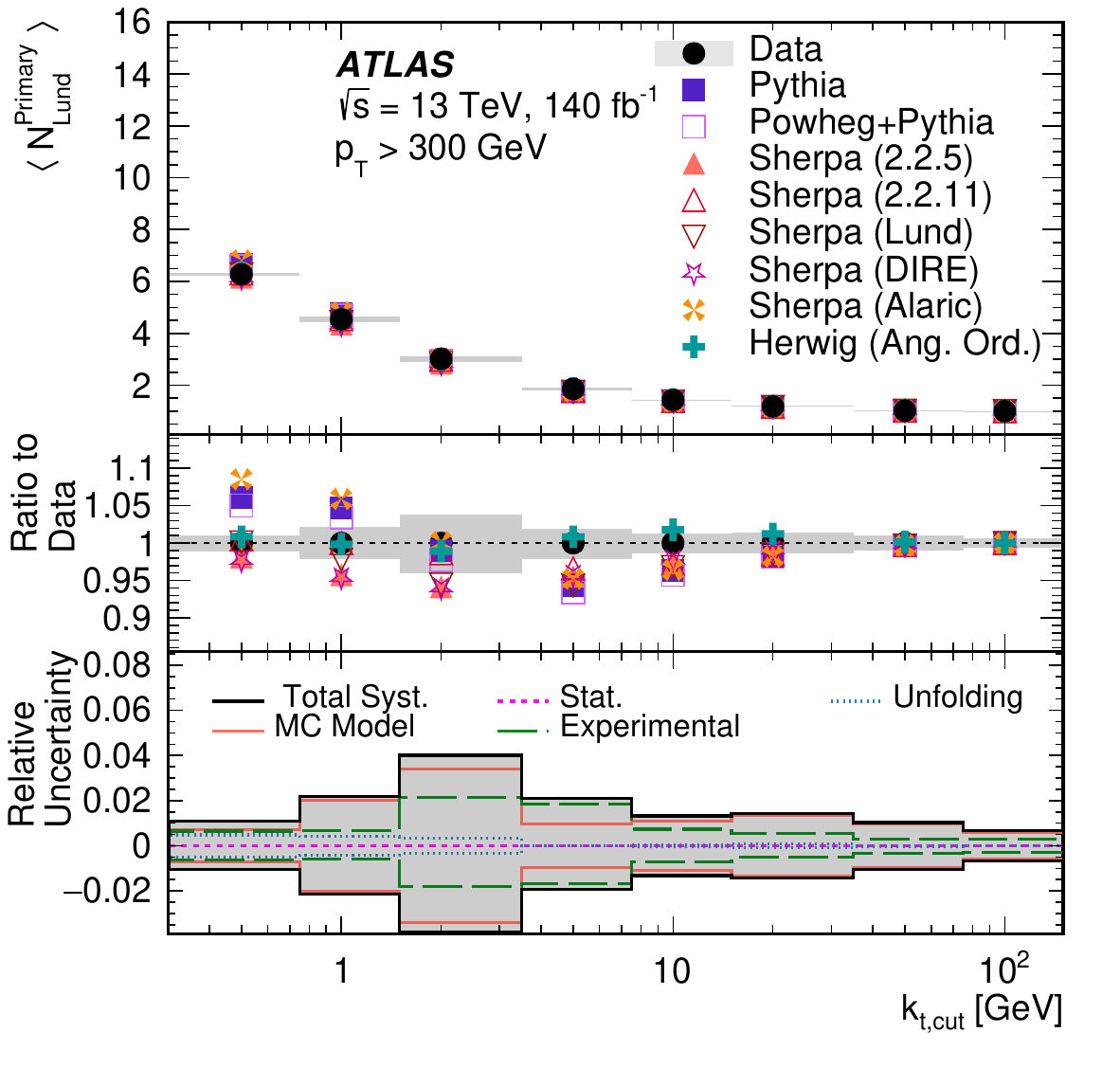}}
\subfigure[]{\includegraphics[width=0.37\textwidth, trim=5 9 15 5, clip]{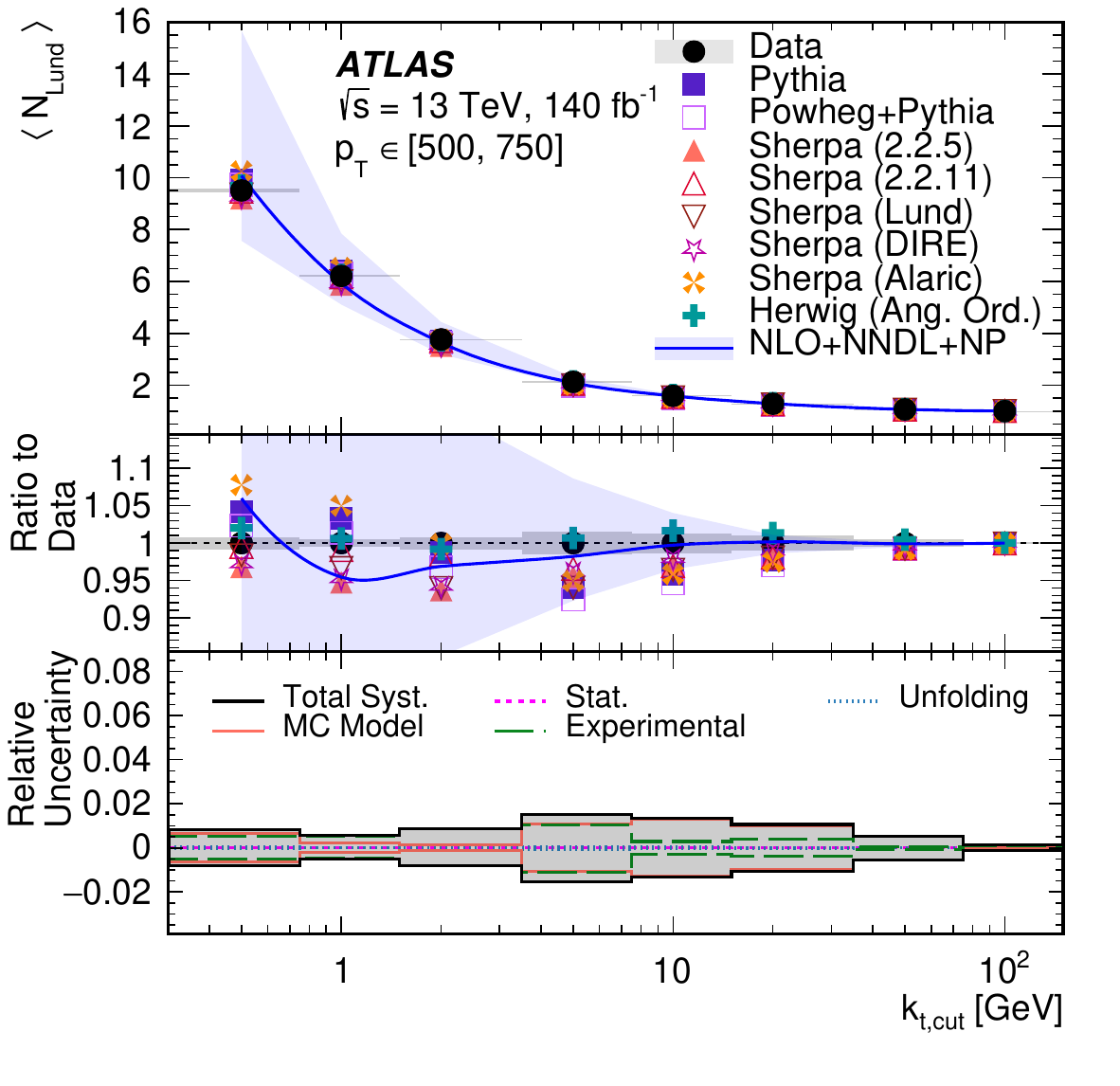}\label{fig:results:avg:c}}
\subfigure[]{\includegraphics[width=0.37\textwidth, trim=5 9 15 5, clip]{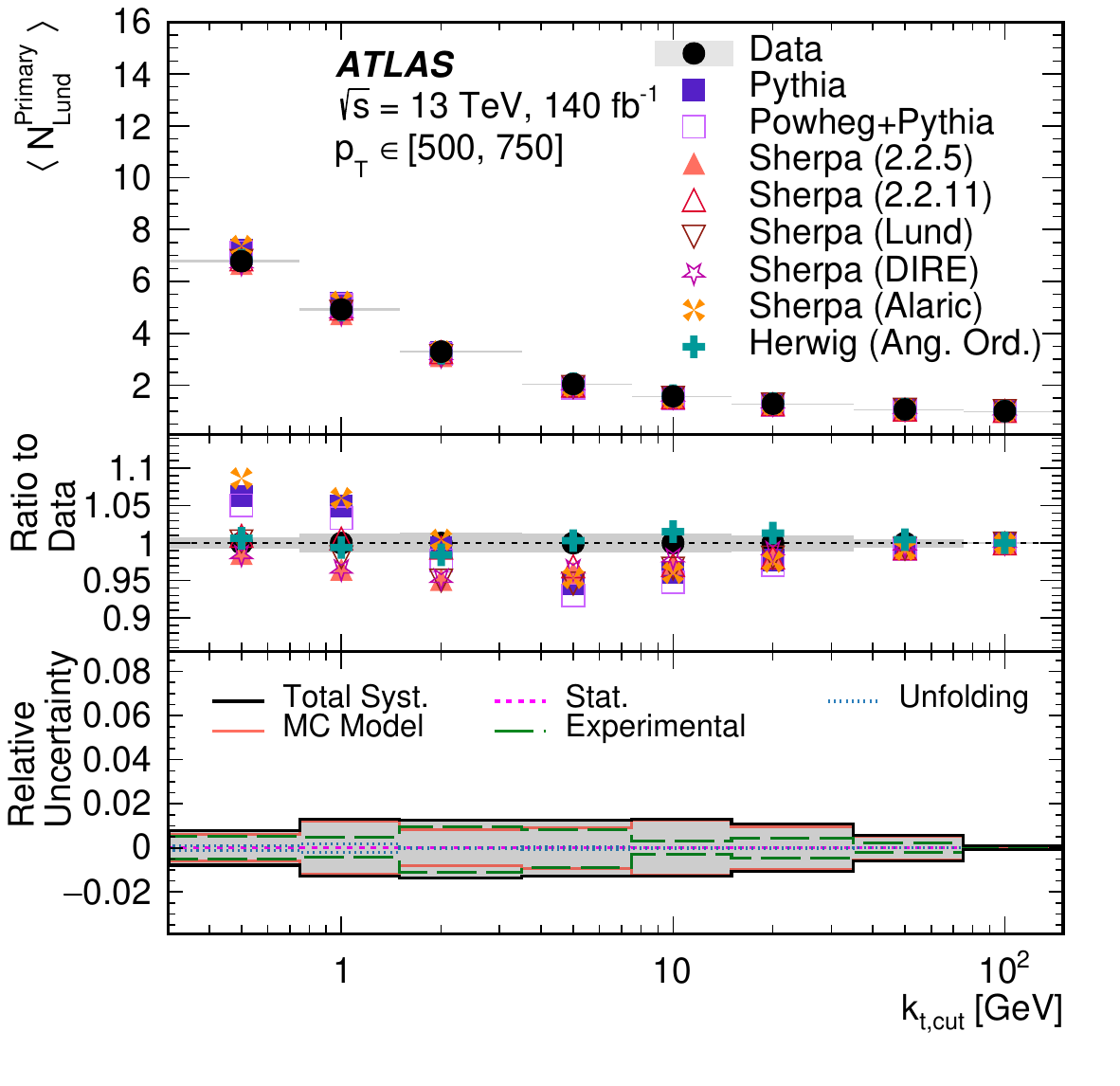}}
\subfigure[]{\includegraphics[width=0.37\textwidth, trim=5 9 15 5, clip]{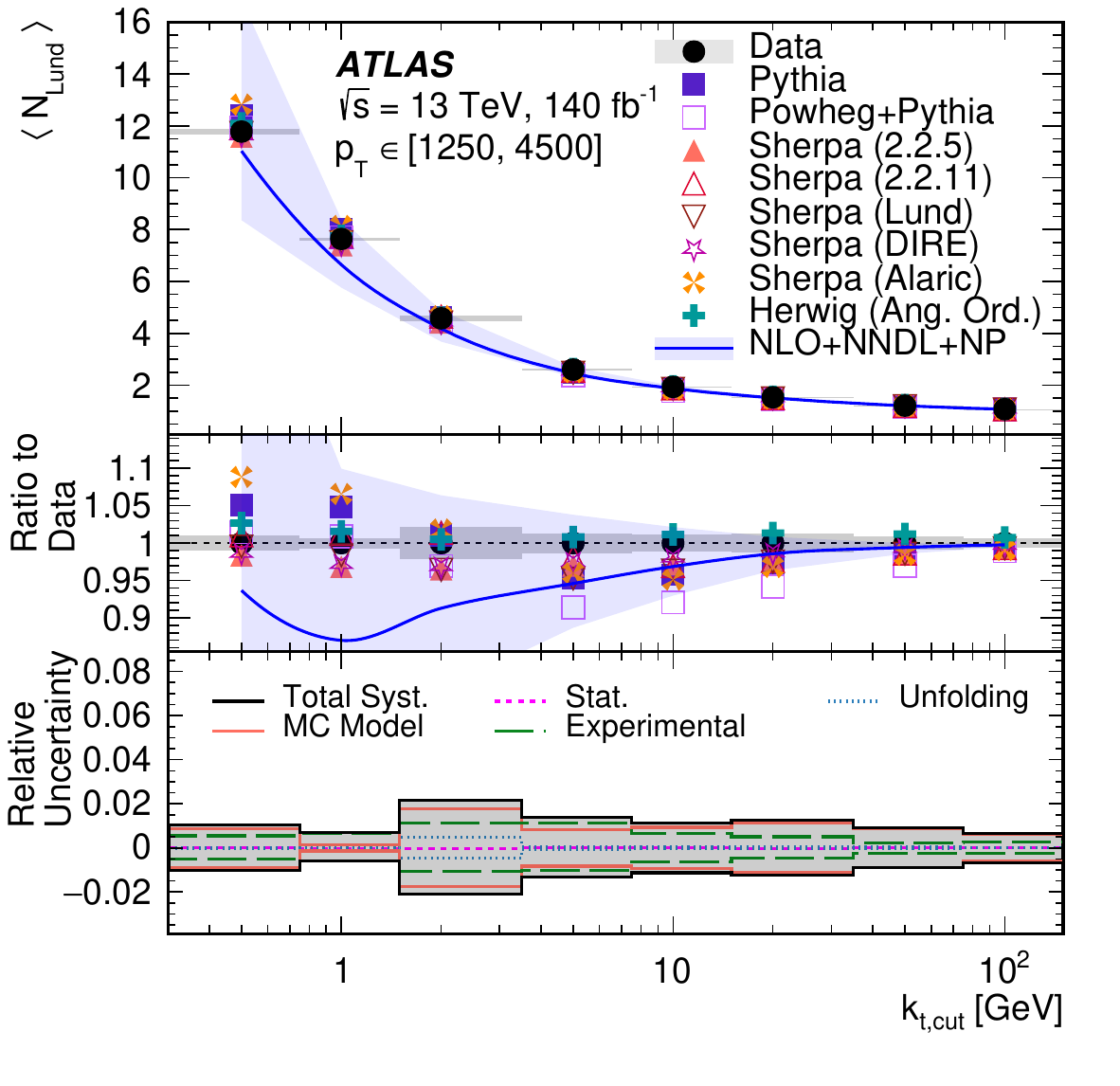}\label{fig:results:avg:e}}
\subfigure[]{\includegraphics[width=0.37\textwidth, trim=5 9 15 5, clip]{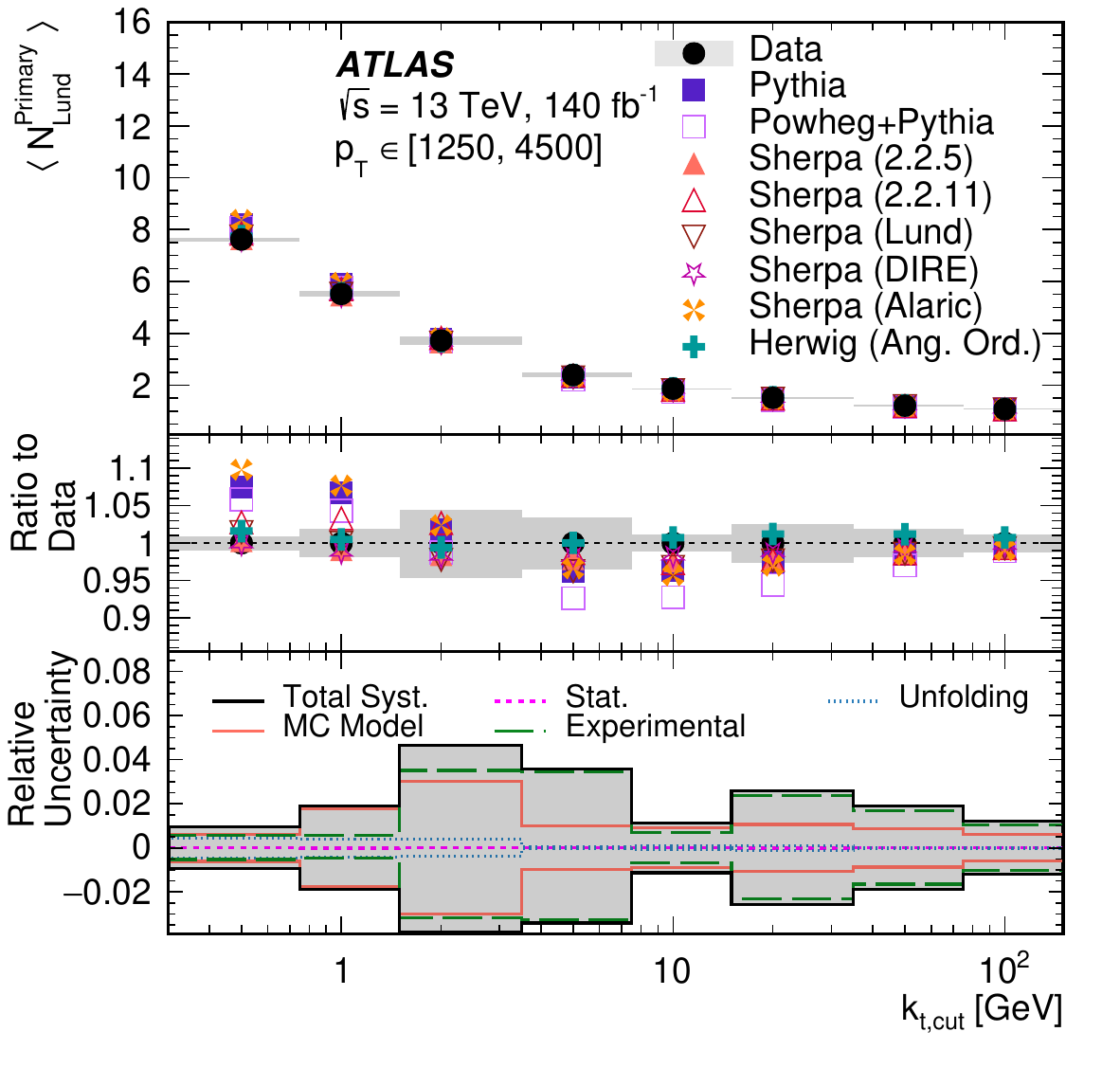}}
\caption{
(left) \AvgNLund and (right) \AvgNPrimary are shown as a function of the emission \kt requirement, $k_{t\text{,cut}}$.
The unfolded data are compared with several MC predictions in (a,b) an inclusive \pt bin above 300~\GeV, (c,d) a \pt bin between 500~\GeV and 750~\GeV and (e,f) a \pt bin between 1250~\GeV and 4500~\GeV.
The \AvgNLund distribution is also compared with an analytic NLO+NNDL+NP prediction with additional non-perturbative corrections, depicted as a solid line, provided by the authors of Ref.~\cite{Medves:2022uii}.
The total uncertainty on the data and the NLO+NNDL+NP prediction are indicated as shaded regions.
The middle panel shows a ratio of the predictions to the measured data, and the bottom panel summarizes the various systematic uncertainties in each bin.
}
\label{fig:results:avg}
\end{figure}

The distributions of \AvgNLund and \AvgNPrimary are shown in Figure~\ref{fig:results:avg} as a function of the emission $\kt$ requirement, for jets with \pt above 300~\GeV.
The agreement between the unfolded data and MC models is qualitatively similar for both observables.
The uncertainty on the measured average multiplicity values as a function of emission \kt is between $0.5-2\%$ for \AvgNLund and $0.5-4\%$ for \AvgNPrimary, increasing with jet \pt.
The level of precision for \AvgNLund is sufficient, for instance, to distinguish the next-to-double-logarithm (NDL) and NNDL predictions of Ref.~\cite{Medves:2022uii}.

Regions of the observable sensitive to non-perturbative effects can be determined by studying the relative separation of the \SHERPA predictions with cluster- and string-based hadronisation models~\cite{STDM-2018-57}: they disagree for $\kt\leq5~\GeV$ and are compatible elsewhere.
The \PYTHIA and \POWPY PSMCs are found to underestimate the measured \AvgNLund data in the perturbative region.
The \SHERPA \DIRE model, which includes some treatment of higher-order splitting functions, performs best among the \SHERPA setups here, but also underestimates the average value in data.
The \HERWIG sample has the best overall agreement for both observables, and is found to be compatible with the unfolded data's uncertainty throughout both the perturbative and non-perturbative regions of the measurement.
The \PYTHIA, \POWPY and \SHERPA \ALARIC samples overestimate the average multiplicity at low \kt ($<1~\GeV$), although the \SHERPA[2.2.5] Lund sample shows good agreement in this region.

The analytic prediction from Ref.~\cite{Medves:2022uii} at NLO matched to NNDL resummation (labelled `NLO+NNDL+NP') was provided by the authors, and is also compared with the \AvgNLund values shown as a function of \kt in Figure~\ref{fig:results:avg:a},~\ref{fig:results:avg:c}, and \ref{fig:results:avg:e}.
This prediction includes non-perturbative corrections computed by taking the ratio of the prediction in MC simulations at hadron-level with multi-parton interactions, and at parton-level without multi-parton interactions, following the prescription in Ref.~\cite{Lifson:2020gua}.
This introduces an additional source of theoretical uncertainty that accounts for the experimental use of charged particles.
Both the non-perturbative correction and its associated uncertainty are small for $\kt>2~\GeV$, but grow rapidly for smaller values where non-perturbative effects become more important.

Agreement between the central value of this prediction and the unfolded data is good in the region where perturbative effects are most relevant ($\kt\geq5~\GeV$), and is competitive with the best PSMC models in the lowest bins of jet \pt.
The uncertainty on the prediction, dominated by the contribution from theoretical scale variations for emission $\kt\geq5~\GeV$, covers the difference between the prediction's central value and the measured data.
As the jet \pt increases, the central value of the NLO+NNDL+NP prediction increasingly underestimates the \AvgNLund distribution; this is most pronounced for low \kt requirements ($\kt\leq10~\GeV$).
In this region, while they are still compatible within the theoretical uncertainty, the difference between the measured data and the prediction's central value approaches 10\%: this is larger than the observed disagreement between the measured data and PSMC samples.

\FloatBarrier

\section{Summary}

In summary, this Letter presents a differential cross-section measurement of Lund subjet multiplicities in dijet events with 140~\ifb~of $\sqrt{s}=13$~\TeV\ $pp$ collision data collected with the ATLAS detector at the Large Hadron Collider.
Multiplicities are calculated by reclustering the jets' constituents with the C/A algorithm, then counting the emissions above a specified \kt requirement in the full clustering tree or its primary branch.
These observables are reconstructed using ID tracks within jets to probe finer angular scales.
Acceptance and detector effects are accounted for using an iterative Bayesian unfolding procedure.

The measurement is made for increasing requirements on the subjet \kt, and the average multiplicity value for each \kt requirement is also extracted for comparison with PSMC and analytic predictions.
Out of many PSMC setups that were studied, a \HERWIG sample generated with the default angle-ordered PS is found to agree best overall with the measured data for both the multiplicity distributions and their extracted average values.
Several recent \SHERPA setups describe the differential multiplicity distributions best when more non-perturbative emissions are allowed ($\kt<2~\GeV$).
Other PSMC models often fail to accurately describe the measured data, indicating the utility of this measurement as an input to future parton shower developments.

The resummed analytic prediction presented in Ref.~\cite{Medves:2022uii} was found to be in good agreement with the measured data in the perturbative region of the measurement ($\kt>2~\GeV$), particularly for jets with low \pt (\emph{e.g.} between 300~\GeV and 500~\GeV).
In this region, the performance of this prediction matches that of the best PSMC models that were studied.
As the jet \pt increases, the performance of this prediction deteriorates, and it is found to underestimate the measured data in the highest \pt bins.
This illustrates the important role of JSS measurements targetting such effects at the LHC, where access to high-energy jets provides a complement to existing fragmentation measurements made at lower-energy facilities such as LEP and HERA.

A Rivet routine is available for this measurement~\cite{Buckley:2010ar}, and the measured data points have been made publicly available~\cite{hepdata.138878} as an input to future parton shower developments and other studies probing fundamental properties of QCD and jet formation at the TeV-scale.

% End of text imported from the .//sections/body.tex input file

%
\section*{Acknowledgements}
%

% The next lines are included from the .//acknowledgements/Acknowledgements.tex input file
%
%

%
%

We thank CERN for the very successful operation of the LHC and its injectors, as well as the support staff at
CERN and at our institutions worldwide without whom ATLAS could not be operated efficiently.

The crucial computing support from all WLCG partners is acknowledged gratefully, in particular from CERN, the ATLAS Tier-1 facilities at TRIUMF/SFU (Canada), NDGF (Denmark, Norway, Sweden), CC-IN2P3 (France), KIT/GridKA (Germany), INFN-CNAF (Italy), NL-T1 (Netherlands), PIC (Spain), RAL (UK) and BNL (USA), the Tier-2 facilities worldwide and large non-WLCG resource providers. Major contributors of computing resources are listed in Ref.~\cite{ATL-SOFT-PUB-2023-001}.

We gratefully acknowledge the support of ANPCyT, Argentina; YerPhI, Armenia; ARC, Australia; BMWFW and FWF, Austria; ANAS, Azerbaijan; CNPq and FAPESP, Brazil; NSERC, NRC and CFI, Canada; CERN; ANID, Chile; CAS, MOST and NSFC, China; Minciencias, Colombia; MEYS CR, Czech Republic; DNRF and DNSRC, Denmark; IN2P3-CNRS and CEA-DRF/IRFU, France; SRNSFG, Georgia; BMBF, HGF and MPG, Germany; GSRI, Greece; RGC and Hong Kong SAR, China; ISF and Benoziyo Center, Israel; INFN, Italy; MEXT and JSPS, Japan; CNRST, Morocco; NWO, Netherlands; RCN, Norway; MEiN, Poland; FCT, Portugal; MNE/IFA, Romania; MESTD, Serbia; MSSR, Slovakia; ARIS and MVZI, Slovenia; DSI/NRF, South Africa; MICIU/AEI, Spain; SRC and Wallenberg Foundation, Sweden; SERI, SNSF and Cantons of Bern and Geneva, Switzerland; NSTC, Taipei; TENMAK, T\"urkiye; STFC/UKRI, United Kingdom; DOE and NSF, United States of America.

Individual groups and members have received support from BCKDF, CANARIE, CRC and DRAC, Canada; CERN-CZ, PRIMUS 21/SCI/017 and UNCE SCI/013, Czech Republic; COST, ERC, ERDF, Horizon 2020, ICSC-NextGenerationEU and Marie Sk{\l}odowska-Curie Actions, European Union; Investissements d'Avenir Labex, Investissements d'Avenir Idex and ANR, France; DFG and AvH Foundation, Germany; Herakleitos, Thales and Aristeia programmes co-financed by EU-ESF and the Greek NSRF, Greece; BSF-NSF and MINERVA, Israel; Norwegian Financial Mechanism 2014-2021, Norway; NCN and NAWA, Poland; La Caixa Banking Foundation, CERCA Programme Generalitat de Catalunya and PROMETEO and GenT Programmes Generalitat Valenciana, Spain; G\"{o}ran Gustafssons Stiftelse, Sweden; The Royal Society and Leverhulme Trust, United Kingdom.

In addition, individual members wish to acknowledge support from CERN: European Organization for Nuclear Research (CERN PJAS); Chile: Agencia Nacional de Investigaci\'on y Desarrollo (FONDECYT 1190886, FONDECYT 1210400, FONDECYT 1230812, FONDECYT 1230987); China: Chinese Ministry of Science and Technology (MOST-2023YFA1605700), National Natural Science Foundation of China (NSFC - 12175119, NSFC 12275265, NSFC-12075060); Czech Republic: PRIMUS Research Programme (PRIMUS/21/SCI/017); EU: H2020 European Research Council (ERC - 101002463); European Union: European Research Council (ERC - 948254, ERC 101089007), Horizon 2020 Framework Programme (MUCCA - CHIST-ERA-19-XAI-00), European Union, Future Artificial Intelligence Research (FAIR-NextGenerationEU PE00000013), Italian Center for High Performance Computing, Big Data and Quantum Computing (ICSC, NextGenerationEU); France: Agence Nationale de la Recherche (ANR-20-CE31-0013, ANR-21-CE31-0013, ANR-21-CE31-0022, ANR-22-EDIR-0002), Investissements d'Avenir Labex (ANR-11-LABX-0012); Germany: Baden-Württemberg Stiftung (BW Stiftung-Postdoc Eliteprogramme), Deutsche Forschungsgemeinschaft (DFG - 469666862, DFG - CR 312/5-2); Italy: Istituto Nazionale di Fisica Nucleare (ICSC, NextGenerationEU), Ministero dell'Università e della Ricerca (PRIN - 20223N7F8K - PNRR M4.C2.1.1); Japan: Japan Society for the Promotion of Science (JSPS KAKENHI JP21H05085, JSPS KAKENHI JP22H01227, JSPS KAKENHI JP22H04944, JSPS KAKENHI JP22KK0227); Netherlands: Netherlands Organisation for Scientific Research (NWO Veni 2020 - VI.Veni.202.179); Norway: Research Council of Norway (RCN-314472); Poland: Polish National Agency for Academic Exchange (PPN/PPO/2020/1/00002/U/00001), Polish National Science Centre (NCN 2021/42/E/ST2/00350, NCN OPUS nr 2022/47/B/ST2/03059, NCN UMO-2019/34/E/ST2/00393, UMO-2020/37/B/ST2/01043, UMO-2021/40/C/ST2/00187, UMO-2022/47/O/ST2/00148); Slovenia: Slovenian Research Agency (ARIS grant J1-3010); Spain: BBVA Foundation (LEO22-1-603), Generalitat Valenciana (Artemisa, FEDER, IDIFEDER/2018/048), Ministry of Science and Innovation (MCIN \& NextGenEU PCI2022-135018-2, MICIN \& FEDER PID2021-125273NB, RYC2019-028510-I, RYC2020-030254-I, RYC2021-031273-I, RYC2022-038164-I), PROMETEO and GenT Programmes Generalitat Valenciana (CIDEGENT/2019/023, CIDEGENT/2019/027); Sweden: Swedish Research Council (VR 2018-00482, VR 2022-03845, VR 2022-04683, VR grant 2021-03651), Knut and Alice Wallenberg Foundation (KAW 2017.0100, KAW 2018.0157, KAW 2018.0458, KAW 2019.0447, KAW 2022.0358); Switzerland: Swiss National Science Foundation (SNSF - PCEFP2\_194658); United Kingdom: Leverhulme Trust (Leverhulme Trust RPG-2020-004), Royal Society (NIF-R1-231091); United States of America: U.S. Department of Energy (ECA DE-AC02-76SF00515), Neubauer Family Foundation.

%
%

% End of text imported from the .//acknowledgements/Acknowledgements.tex input file

%
%
%
%
%
%

%

%
%

%
%
\printbibliography
\clearpage
% ATLAS Collaboration author list
% Reference date of STDM-2023-07 is 2023-12-08
% Author list last updated on date 04-SEP-24
% Data extracted on 04-Sep-2024 for paper reference STDM-2023-07
% at 10:56am
 
\begin{flushleft}
\hypersetup{urlcolor=black}
{\Large The ATLAS Collaboration}

\bigskip

\AtlasOrcid[0000-0002-6665-4934]{G.~Aad}$^\textrm{\scriptsize 103}$,
\AtlasOrcid[0000-0001-7616-1554]{E.~Aakvaag}$^\textrm{\scriptsize 16}$,
\AtlasOrcid[0000-0002-5888-2734]{B.~Abbott}$^\textrm{\scriptsize 121}$,
\AtlasOrcid[0000-0002-1002-1652]{K.~Abeling}$^\textrm{\scriptsize 55}$,
\AtlasOrcid[0000-0001-5763-2760]{N.J.~Abicht}$^\textrm{\scriptsize 49}$,
\AtlasOrcid[0000-0002-8496-9294]{S.H.~Abidi}$^\textrm{\scriptsize 29}$,
\AtlasOrcid[0009-0003-6578-220X]{M.~Aboelela}$^\textrm{\scriptsize 44}$,
\AtlasOrcid[0000-0002-9987-2292]{A.~Aboulhorma}$^\textrm{\scriptsize 35e}$,
\AtlasOrcid[0000-0001-5329-6640]{H.~Abramowicz}$^\textrm{\scriptsize 152}$,
\AtlasOrcid[0000-0002-1599-2896]{H.~Abreu}$^\textrm{\scriptsize 151}$
% ...
\bigskip
\\

% ...
$^{16}$Department for Physics and Technology, University of Bergen, Bergen; Norway.\\
% ...
$^{29}$Physics Department, Brookhaven National Laboratory, Upton NY; United States of America.\\
% ...
$^{35}$$^{(a)}$Facult\'e des Sciences Ain Chock, Universit\'e Hassan II de Casablanca;$^{(b)}$Facult\'{e} des Sciences, Universit\'{e} Ibn-Tofail, K\'{e}nitra;$^{(c)}$Facult\'e des Sciences Semlalia, Universit\'e Cadi Ayyad, LPHEA-Marrakech;$^{(d)}$LPMR, Facult\'e des Sciences, Universit\'e Mohamed Premier, Oujda;$^{(e)}$Facult\'e des sciences, Universit\'e Mohammed V, Rabat;$^{(f)}$Institute of Applied Physics, Mohammed VI Polytechnic University, Ben Guerir; Morocco.\\
% ...
$^{44}$Physics Department, Southern Methodist University, Dallas TX; United States of America.\\
% ...
$^{49}$Fakult\"{a}t Physik , Technische Universit{\"a}t Dortmund, Dortmund; Germany.\\
% ...
$^{55}$II. Physikalisches Institut, Georg-August-Universit\"{a}t G\"ottingen, G\"ottingen; Germany.\\
% ...
$^{103}$CPPM, Aix-Marseille Universit\'e, CNRS/IN2P3, Marseille; France.\\
% ...
$^{121}$Homer L. Dodge Department of Physics and Astronomy, University of Oklahoma, Norman OK; United States of America.\\
% ...
$^{151}$Department of Physics, Technion, Israel Institute of Technology, Haifa; Israel.\\
$^{152}$Raymond and Beverly Sackler School of Physics and Astronomy, Tel Aviv University, Tel Aviv; Israel.\\
% ...
$^{a}$ Also Affiliated with an institute covered by a cooperation agreement with CERN.\\
$^{b}$ Also at An-Najah National University, Nablus; Palestine.\\
$^{c}$ Also at Borough of Manhattan Community College, City University of New York, New York NY; United States of America.\\
$^{d}$ Also at Center for High Energy Physics, Peking University; China.\\
$^{e}$ Also at Center for Interdisciplinary Research and Innovation (CIRI-AUTH), Thessaloniki; Greece.\\
$^{f}$ Also at Centro Studi e Ricerche Enrico Fermi; Italy.\\
% ...

\end{flushleft}

% Created with Glance <Atlas.Glance@cern.ch>

%

%
%
%
%
%

%
%
%
%

\end{document}